 \documentclass[final,5p,times,twocolumn]{elsarticle}

\usepackage[T1]{fontenc} 

\usepackage{amssymb}

\usepackage{textcomp}
\usepackage{xspace}
\usepackage{xcolor}
\usepackage{txfonts}
\usepackage{booktabs}

\usepackage{hyperref}

\journal{Molecular Physics}

\begin{document}

\begin{frontmatter}

\title{Millimeter and submillimeter spectroscopy of methylallene, CH$_3$CHCCH$_2$}

\author[Koeln]{Holger S.~P. M\"uller\corref{cor}} 
\ead{hspm@ph1.uni-koeln.de} 
\cortext[cor]{Corresponding author.} 
\author[Koeln]{Frank Lewen} 
\author[Rennes]{Jean-Claude Guillemin} 
\author[Koeln]{Stephan Schlemmer}

\address[Koeln]{I.~Physikalisches Institut, Universit{\"a}t zu K{\"o}ln, 
  Z{\"u}lpicher Str. 77, 50937 K{\"o}ln, Germany}
\address[Rennes]{Univ Rennes, Ecole Nationale Sup{\'e}rieure de Chimie de Rennes, 
  CNRS, ISCR$-$UMR 6226, 35000 Rennes, France}

\begin{abstract}
Small polycyclic aromatic hydrocarbons and somewhat larger cyano derivatives were detected in the cold dark cloud TMC-1 recently. 
Their formation from smaller hydrocarbons is not well understood, in part because abundances of many species are not known. 
Methylallene, CH$_3$CHCCH$_2$,  may be one of the building blocks, but its rotational spectrum was characterized only to a very limited extent. 
We recorded rotational transitions in the 36$-$501~GHz region to extend the existing line list of methylallene and thus enable searches for the molecule in space. 
Quantum-chemical calculations were carried out to evaluate initial spectroscopic parameters. 
We obtained transition frequencies with $J \le 61$ and $K_a \le 21$ and resolved the internal rotation splitting of the CH$_3$ group at least partially. 
As a result, a full set of distortion parameters up to sixth order along with two octic ones were determined, 
as well as parameters describing the internal rotation of the methyl group. 
The spectroscopic parameters are accurate enough to identify methylallene up to 720~GHz, sufficient for searches even in the warm interstellar medium. 
\\\resizebox{22pc}{!}{\includegraphics{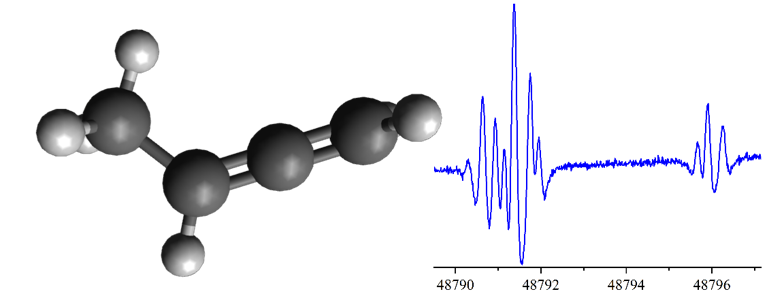}}
\end{abstract}


\end{frontmatter}

\section{Introduction}
\label{intro}

More than 340 different molecules have been detected in the interstellar medium (ISM) 
and in circumstellar shells of late-type stars thus far, see the Molecules in Space 
webpage \cite{mols-in-space} of the Cologne Database for Molecular Spectroscopy, 
CDMS \cite{CDMS_2005,CDMS_2016}, for an up-to-date list. 
A substantial fraction of the recent detections was made in the cold and dark molecular cloud TMC-1. 
They include the polycyclic aromatic hydrocarbon (PAH) molecules indene \cite{indene_det-GBT_2021,indene_etc_det-Pepe_2021} 
and phenalene \cite{phenalene_det_2025} along with several PAH cyano derivatives up to 
cyanocoronene \cite{cyanocoronene_det_2025}, a molecule containing seven fused benzene rings. 
The abundances were interpreted in favor of a formation of these molecules in TMC-1 from 
smaller hydrocarbons (bottom-up chemistry) rather than from a break-up of even larger molecules 
(top-down chemistry) \cite{bottom-up_2024}. 
Several small to moderately sized hydrocarbons were detected in space, in earlier years often very unsaturated ones 
with more C than H atoms, in part because they often have substantial dipole moments. 
More saturated hydrocarbons usually have small or even zero dipole moments, 
making it difficult or impossible to detect them by means of radio astronomy. However, chemical aspects are very important also. 
Unsaturated molecules, often very unsaturated ones, are very abundant in cold dark molecular clouds 
whereas (fairly) saturated molecules dominate in the warm parts of star-forming regions. 
There is a fairly common misconception that abundances of molecules in space would be governed by their relative energies. 
For example among the C$_3$H$_4$O isomers, propadienone (H$_2$CCCO) has been elusive until very recently \cite{H2C3O_det_2026} 
even though it is lower in energy than propynal and much lower than cyclopropenone.  
Shingledecker et al. \cite{H2C3O_kinetics_2019} showed that the reaction between propadienone and atomic hydrogen is barrierless, 
making it difficult to accumulate propadienone in space. 
But kinetics does not only govern astrochemistry, it is also dominant under more ambient conditions. 
For example the primary products in the reaction of Cl atoms with NO$_2$ at room temperature with an excess of N$_2$ are ClONO and ClNO$_2$ 
with ClONO initially four times as abundant as ClNO$_2$ despite the latter being $\sim$10~kcal/mol lower in energy \cite{Cl+NO2_1978}.

Propene, C$_3$H$_6$, also known as (aka) propylene, was detected nearly 20 years ago in TMC-1 \cite{propene_det_2007}, 
was later found in several other cold clouds \cite{HCCO_det_and_more_propene_2015,isobutene_rot_det_2023}, 
and was even found in the lukewarm parts surrounding the low-mass protostar IRAS 16292-2422B \cite{propenal_propene_2020}. 
More recently, isobutene, (CH$_3$)$_2$CCH$_2$, was identified in the cold and presumably shocked molecular cloud 
G+0.693$-$0.027 \cite{isobutene_rot_det_2023} in the vicinity of the high-mass star forming region Sagittarius B2(N) 
close to the Galactic center, and 1-butyne, C$_2$H$_5$CCH, aka ethylacetylene, 
was found unambiguously soon thereafter in TMC-1 \cite{butyne_etc_det_2024}.

\begin{figure}
\centering
 \includegraphics[width=6cm,angle=0]{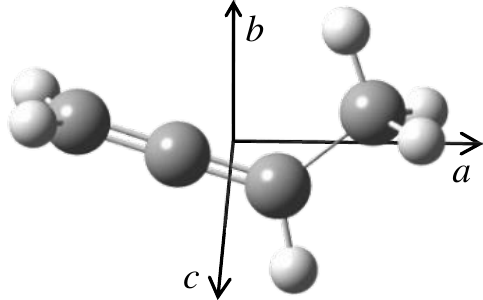}

\caption{Sketch of the methylallene molecule. Carbon atoms are symbolized by gray spheres 
   while hydrogen atoms are indicated by small, light gray spheres. The $a$-axis is in the paper plane  
   and the $b$-axis is rotated slightly toward the viewer.}
\label{mol-bild}
\end{figure}

Methylallene (Figure~\ref{mol-bild}), CH$_3$CHCCH$_2$, aka 1,2-butadiene, is an isomer of 1- and 2-butyne and also of 1,3-butadiene. 
The low-energy \textit{trans}-1,3-butadiene conformer has no permanent dipole moment, such that methylallene may serve as a proxy 
for the former and possibly also for allene, H$_2$CCCH$_2$, which also has no permanent dipole moment. 
It should be pointed out that there is a \textit{gauche} conformer of 1,3-butadiene with a very small dipole moment, 
which was characterized by rotational spectroscopy \cite{g-13BD_rot_2017}, but it is 2.93~kcal/mol (1475~K) higher in energy \cite{g-13BD_UV_egy_2001}. 
While 1,3-butadiene was thought to be an important molecule for the formation of at least some PAH molecules, 
this role was questioned in a recent rebuttal of the detection of 1-cyano-1,3-butadiene, aka 2,4-pentadienenitrile \cite{C5H5N_rebuttal_2025}. 
It appears plausible that methylallene can act as an additional building block for larger hydrocarbons, 
even more so, as the related ethynylallene, aka allenylacetylene or 1,2-pentadien-4-yne is among 
the fairly unsaturated molecules detected in space \cite{allenylacetylene_det_2021}.

The rotational spectrum of methylallene was studied almost 70 years ago in the microwave region by Lide and Mann \cite{methylallene_rot_1957}  
with observations of splittings, caused by the hindered internal rotation of the methyl group, often resolved and with the determination of dipole moment components. 
A marginal extension of the spectrum to $\sim$50~GHz was reported by Ogata \cite{methylallene_isos_rot_1992} in the course of a structure determination of methylallene.

We studied the rotational spectrum of methylallene extensively up to submillimeter wavelengths in order to be able to elucidate its importance in astrochemistry. 
The resulting spectroscopic parameters are deemed to be accurate enough to search for methylallene not only in cold and dark molecular clouds, 
but also in the warmer parts of star-forming regions because of the accurately measured higher frequency transitions.


\section{Experimental details}
\label{exptl}

\subsection{Sample preparation}
\label{syntheses}

The original synthesis \cite{synthese_12BD_1931} was reproduced with some minor modifications in the final step: 
the reflux condenser was maintained at 20$-$25$^{\rm o}$C, and nitrogen was used instead of carbon dioxide for bubbling. 
The crude product thus obtained was purified by trap-to-trap distillation under vacuum (0.1~mbar). 
The high-boiling compounds were condensed in the first trap, immersed in a cooling bath at $-$100$^{\rm o}$C, 
while methylallene was selectively condensed in the second trap, immersed in a liquid nitrogen bath. 
This purification was repeated once. A yield of 68\% was determined by weighing the trapped compound 
and by $^1$H NMR analysis, which revealed the presence of 1$-$2\% ethanol.

\subsection{Spectroscopic measurements}
\label{lab-spec}

Measurements at the Universit{\"a}t zu K{\"o}ln were carried out between 36 and 501~GHz applying two slightly different spectrometer setups. 
One is equipped with two connected 7~m long glass cells in a single path arrangement sealed with Teflon windows and with different Schottky diode detectors. 
An Agilent E8257D synthesizer at fundamental frequencies was used as source for measurements between 35 and 69~GHz, 
while this synthesizer together with a VDI frequency tripler was employed between 71 and 117~GHz. 
Employing a frequency multiplier from RPG instead of the VDI frequency tripler enabled measurements between 143 and 165~GHz. 
The sample pressure was between 2.0 and 3.0~Pa, and the frequency modulation was chosen so that the intensities of stronger lines were slightly below their maximum. 
This setup is similar to the one employed by Ordu et al. \cite{n-BuCN_rot_2012}.

A combination of an active and passive frequency multipliers from the VDI starter kit driven by a Rohde \& Schwarz SMF~100A synthesizer 
was used for spectral recordings in the 344 to 501~GHz region. 
A 5~m long double path cell was equipped with a Teflon window and a retro reflector which rotated the polarisation by 90$^{\rm o}$. 
A Schottky diode was employed as detector. 
The spectrometer was described in detail earlier \cite{OSSO_rot_2015}.

In order to achieve narrow lines, the modulation was lowered for these measurements to the point at which the line-width decrease was still more pronounced than the loss in intensity. 
The sample pressure could be raised to between 5.0 and 7.0~Pa without substantial increase in line-width. 
Frequency modulation was applied throughout; the demodulation at $2f$ causes the lines to appear close to the second derivative of a Gaussian. 
As the rotational spectrum of methylallene is fairly weak even for the strongest lines, only individual transitions or small groups thereof were recorded 
with integration times adjusted to reach good signal-to-noise ratios.

\section{Quantum-chemical calculation}
\label{qcc}

We carried out quantum-chemical calculations at the Regionales Rechenzentrum der Universit{\"a}t 
zu K{\"o}ln (RRZK) using the commercially available program Gaussian~16 \cite{Gaussian16C}. 
We performed B3LYP hybrid density functional calculations \cite{Becke_1993,LYP_1988} 
employing the correlation consistent basis set augmented with diffuse basis functions 
aug-cc-pVTZ \cite{cc-pVXZ_1989}, the basis set is abbreviated here to 3a, and the entire model is described as B3LYP/3a. 
An equilibrium geometry was determined by analytic gradient techniques, the harmonic force field 
by analytic second derivatives, and the anharmonic force field by numerical differentiation 
of the analytically evaluated second derivatives of the energy. 
The main goal of the anharmonic force field calculation was the determination of ground state rotational parameters 
along with equilibrium quartic and sextic centrifugal distortion parameters.

\section{Spectroscopic properties of methylallene and previous investigations}
\label{rot_backgr}

Methylallene is an asymmetric rotor of the prolate type with $\kappa = (2B - A - C)/(A - C)$ 
being with $-$0.9818 quite close to the limiting case of $-$1. 
A model of the molecule is shown in Figure~\ref{mol-bild}. 
The molecule possesses a fairly small $a$-dipole moment component of 0.397~D and an even smaller $b$-component of 0.071~D. 
Please note that these values differ from the reported $0.394 \pm 0.002$~D and $0.070 \pm 0.001$~D \cite{methylallene_rot_1957} 
because a value of 0.709~D was applied for the OCS reference whereas the current best value is 0.71519~D \cite{OCS_dip_1985,OCS_dip_1986}. 
The strongest $a$-type transitions, $R$-branch transitions having $J' - J'' = +1$, are about a factor of 20 stronger 
near 40~GHz than the strongest $b$-type transitions, which are $K_ a = 1 - 0$ $Q$-branch ($J' - J'' = 0$) transitions. 
Other $b$-type transitions as well as $a$-type $Q$-branch transitions are weaker still.

Lide and Mann studied the microwave spectrum of methylallene between 15.9 and 33.1~GHz 
and recorded the $J = 2 - 1$ to $4 - 3$ $a$-type rotational transitions \cite{methylallene_rot_1957}. 
They resolved in many cases the splitting caused by the hindered internal rotation of the CH$_3$ group 
which amounted to a few megahertz in the most favorable cases. 
They derived a barrier height $V_3$ to internal rotation of $556 \pm 2$~cm$^{-1}$ (800~K). 
The splitting between the totally symmetry A component and the doubly degenerate E component 
in the $a$-type $R$-branch transitions is usually less than $\sim$3~MHz, 
whereas it is up to $\sim$50~MHz for $b$-type and for $a$-type $Q$-branch transitions.

In cases in which the asymmetry splitting of a pair of rotational energies with the same $J$ and the same $K_a$ 
is close to the splitting between the A and E levels, mixing may occur between the E components. 
This mixing causes the energy levels to repel each other and results in larger A/E splittings particularly 
noteworthy in the $a$-type $R$-branch transitions. 
An additional effect of the mixing between the two levels is that usually forbidden $x$-type ($\Delta K_a$ and $\Delta K_c$ are 0~mod~2) 
or $c$-type ($\Delta K_a = 1$ mod~2 and $\Delta K_c = 0$ mod~2) transitions borrow some intensity from the strongly allowed transitions. 
The spin-statistics of the CH$_3$ group are such that the A and E components of a given transition 
have the same intensities in the absence of such interactions.

In a study of several methylallene isotopologs in order to determine structural parameters of the molecule, 
Ogata determined $J = 6 - 5$ A symmetry transition frequencies of the main species 
around 48~GHz, but these were soon replaced by present, more accurate measurements \cite{methylallene_isos_rot_1992}.


\begin{figure}
\centering
 \includegraphics[width=8cm,angle=0]{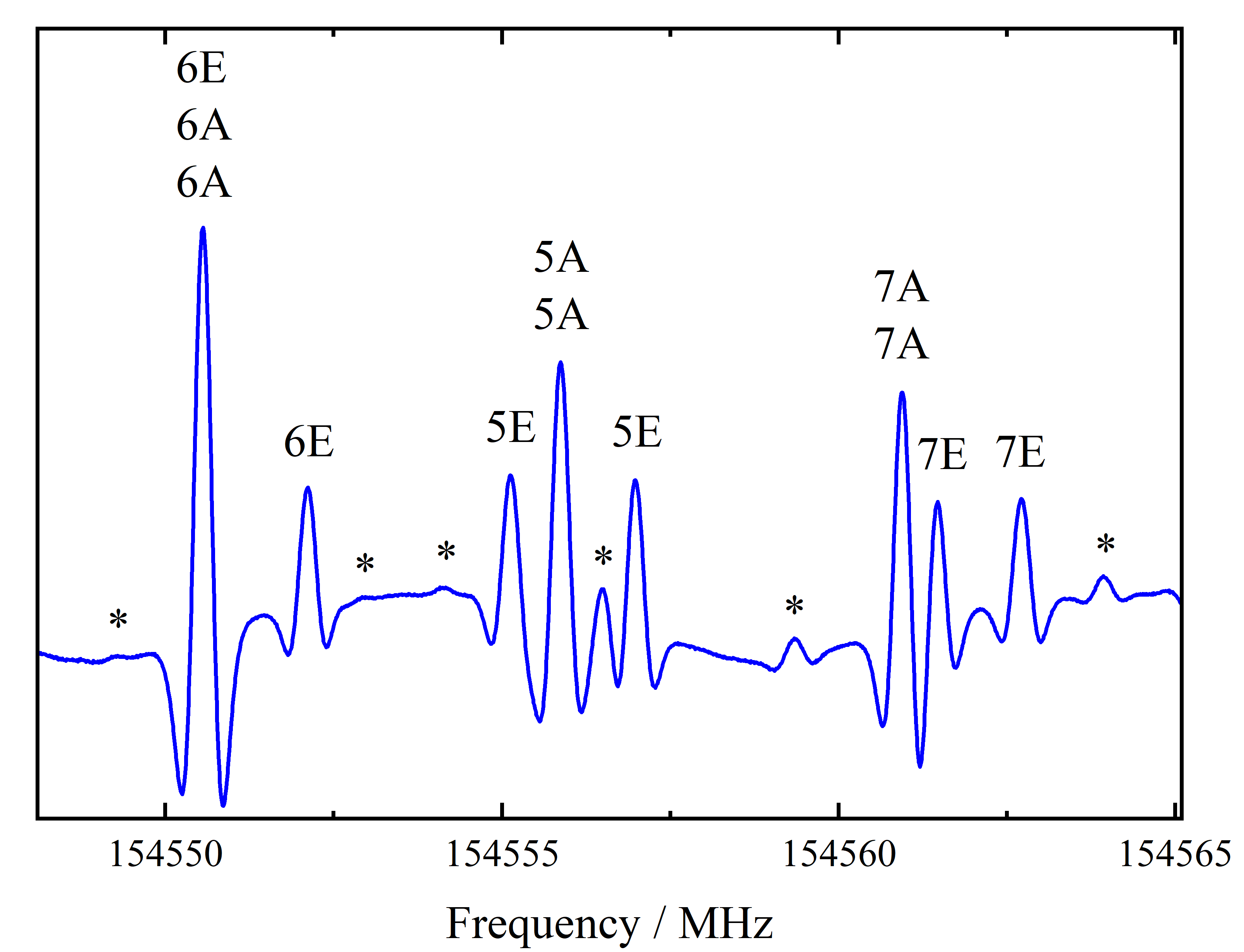}

\caption{Section of the millimeter wave spectrum of methylallene in the region of the $J = 19 - 18$ higher $K_a$ $a$-type transitions. 
   The $K_a$ quantum numbers as well as the methyl internal rotor symmetry labels A and E are indicated. 
   The asymmetry splitting of the two transitions in each higher $K_a$ is unresolved in the A components, while the E components occur in different locations. 
   Several unassigned lines, indicated by asterisks, probably originate from excited vibrational states of methylallene.}
\label{Spektrum_19-18}
\end{figure}

\section{Spectroscopic results}
\label{lab-results}

The rotational spectrum of methylallene was fit and calculated with the Erham program \cite{erham_1997,erham_2012}. 
The low-order parameters describing the internal rotation are $\rho$, $\beta$, and $\epsilon$, which represent 
the length of the $\rho$-vector, its angle with the $a$-axis, and the first order tunneling parameter, respectively. 
Strictly speaking, $\epsilon$ should read $\epsilon _1$, but since we do not utilize any higher order tunneling parameters, we omitted the subscript. 
Furthermore, we prefer shorter notations $\epsilon _K$, $\epsilon _J$, $\epsilon _2$, etc. for $[A - (B + C)/2]$, $[(B + C)/2]$, $[(B - C)/4]$, etc., respectively, 
with one reason being that these notations are more common in other programs.

As the previous line list was very limited, quartic centrifugal distortion parameters from a 
quantum-chemical calculation described in Sect.~\ref{qcc} were included into the parameter set 
and only released if this resulted in a sufficient improvement of the weighted rms.

In the first measurement rounds, $a$-type $R$-branch transitions were recorded in the 36$-$69~GHz region. 
As can be seen in Figure~\ref{Spektrum_19-18} for somewhat higher frequencies, the internal rotation patterns of transitions with higher $K_a$ are distinct. 
The asymmetry splitting between the A components is often unresolved, and the E components may occur on either side or may sometimes be blended 
with the two A components, as in the case of $K_a = 6$ in Figure~\ref{Spektrum_19-18}. 
There are also several unassigned lines in that figure which are likely caused by vibrational satellites, 
which sometimes affect the positions of the $a$-type $R$-branch transitions. 
Interestingly, it appeared as if $b$-type lines were less affected by blending in spite of their lower intensities 
as long as they were not close to frequency regions with $a$-type $R$-branch transitions. 
These measurements of $a$-type $R$-branch transitions constrained $B$, $C$, $D_J$, $D_{JK}$ and $d_1$ quite well along with $\rho$, $\beta$, and $\epsilon$. 
Even $A$ was quite well constrained with an uncertainty of 0.63~MHz after 150~MHz initially \cite{methylallene_rot_1957} 
and 25~MHz later \cite{methylallene_isos_rot_1992}. 
Therefore, we searched for the relatively strong $^rQ_0$ $b$-type transitions. 
The superscript indicated the change from $K_a''$ to $K_a'$, and the subscript indicates that $K_a''$ is 0. 
The E symmetry components were found well within the uncertainties while the A symmetry components appeared just within uncertainties.

Subsequently, further $b$-type transitions were recorded, first in the 36$-$69~GHz region and later in the 71$-$117~GHz and 143$-$165~GHz regions. 
These covered $^pR_1$ and $^rR_0$ transitions and went in $K_a$ up to the $^rP_5$ and $^rP_6$ transitions, extending to $J = 64$ and $K_a = 7$. 
Generally, we tried to record as many transitions belonging to a particular series as possible with reasonable integration times. 
In order to constrain the asymmetry splitting and the internal rotation parameters further, $a$-type $Q$-branch transitions with $K_a = 1$ to 3 were recorded.

In all frequency regions, including the 344$-$501~GHz region, we recorded $a$-type $R$-branch transitions. 
We reached the $K_a = J''$ levels up to $J'' = 17$, covering $J'' = 4$ to 13 and 17 to 19, while we accessed $K_a$ up to 21 in the submillimeter region. 
Extrapolation of the Hamiltonian model into the highest frequency region was very good, in particular for 
transitions with relatively low $J$ ($\lesssim 50$) and $K_a$ up to about 10 because the $b$-type transitions recorded at lower frequencies extended to $J = 64$. 
Slightly more pronounced deviations occured at even higher $J$ and $K_a$, which mainly involved the internal rotation patterns.


\begin{table*}
\begin{center}
\caption{Experimental spectroscopic parameters of methylallene in comparison to values calculated at the B3LYP/3a level 
     and previous experimental rotational and distortion parameters.}
{
\begin{tabular}[t]{lr@{}lr@{}lr@{}lr@{}l}
\toprule 
Parameter & \multicolumn{2}{c}{Experimental} & \multicolumn{2}{c}{B3LYP/3a\textsuperscript{a}} & 
  \multicolumn{2}{c}{Previous \cite{methylallene_rot_1957}} & \multicolumn{2}{c}{Previous \cite{methylallene_isos_rot_1992}} \\
\midrule 
$A$                                   &  33997&.87067~(18)   &  34735&.5   &  33860&.~(150) &  34021&.~(25)    \\
$B$                                   &   4201&.2822272~(99) &   4181&.1   &   4201&.26~(1) &   4201&.301~(10) \\
$C$                                   &   3928&.1001862~(97) &   3919&.5   &   3928&.11~(1) &   3928&.122~(9)  \\
$D_K \times 10^3$                     &   1163&.4093~(78)    &   1146&.8   &       &        &       &          \\
$D_{JK} \times 10^3$                  &  $-$55&.10624~(24)   &  $-$52&.42  &  $-$54&.       &  $-$53&.6~(2)    \\
$D_J \times 10^3$                     &      1&.7802449~(24) &      1&.636 &      1&.2      &      1&.66~(9)   \\
$d_1 \times 10^6$                     & $-$356&.3933~(42)    & $-$315&.1   &       &        &       &          \\
$d_2 \times 10^6$                     &   $-$8&.85211~(98)   &   $-$7&.096 &       &        &       &          \\
$H_K \times 10^6$                     &     92&.98~(11)      &     91&.40  &       &        &       &          \\
$H_{KJ} \times 10^6$                  &   $-$2&.24855~(55)   &   $-$2&.304 &       &        &       &          \\
$H_{JK} \times 10^9$                  & $-$152&.03~(17)      & $-$124&.4   &       &        &       &          \\
$H_J \times 10^{9}$                   &      4&.6398~(23)    &      3&.873 &       &        &       &          \\
$h_1 \times 10^{12}$                  &   1830&.4~(27)       &   1449&.    &       &        &       &          \\
$h_2 \times 10^{12}$                  &    158&.79~(87)      &    117&.8   &       &        &       &          \\
$h_3 \times 10^{12}$                  &     15&.05~(28)      &     10&.59  &       &        &       &          \\
$L_{JJK} \times 10^{15}$              &    402&.~(37)        &       &     &       &        &       &          \\
$l_1 \times 10^{15}$                  &  $-$12&.95~(52)      &       &     &       &        &       &          \\
$\rho$\textsuperscript{b}             &      0&.165597~(32)  &       &     &       &        &       &          \\
$\beta$\textsuperscript{c}            &      5&.9311~(76)    &       &     &       &        &       &          \\
$\epsilon$                            &  $-$49&.999~(13)     &       &     &       &        &       &          \\
$\epsilon _K \times 10^3$             &     22&.03~(75)      &       &     &       &        &       &          \\
$\epsilon _J \times 10^3$             &   $-$1&.773~(41)     &       &     &       &        &       &          \\
$\epsilon _2 \times 10^3$             &   $-$0&.450~(20)     &       &     &       &        &       &          \\
$\epsilon _{KK} \times 10^6$          &  $-$49&.1~(54)       &       &     &       &        &       &          \\
$\epsilon _{JK} \times 10^6$          &   $-$0&.436~(89)     &       &     &       &        &       &          \\
$\epsilon _{JJ} \times 10^6$          &      0&.1711~(75)    &       &     &       &        &       &          \\
$\epsilon _{2K} \times 10^6$          &  $-$16&.45~(39)      &       &     &       &        &       &          \\
$\epsilon _{2J} \times 10^9$          &     89&.0~(37)       &       &     &       &        &       &          \\
$\epsilon _{2KK} \times 10^9$         &     67&.0~(43)       &       &     &       &        &       &          \\
$G _a \times 10^3$                    & $-$320&.2~(53)       &       &     &       &        &       &          \\
$G _{aK} \times 10^3$                 &      1&.198~(67)     &       &     &       &        &       &          \\
$G _{2a} \times 10^6$                 &     15&.8~(27)       &       &     &       &        &       &          \\
$G _{aJK} \times 10^6$                &   $-$0&.176~(11)     &       &     &       &        &       &          \\
$J_{\rm max}$\textsuperscript{b}      &     64&              &       &     &       &        &       &          \\
$K_{a,{\rm max}}$\textsuperscript{b}  &     21&              &       &     &       &        &       &          \\
No. transitions\textsuperscript{b}    &   1137&              &       &     &       &        &       &          \\
No. lines\textsuperscript{b}          &    961&              &       &     &       &        &       &          \\
rms                                   &      0&.0119         &       &     &       &        &       &          \\
wrms\textsuperscript{b}               &      0&.823          &       &     &       &        &       &          \\
\bottomrule 
\end{tabular}
}
\end{center}
 
Notes: Watson's $S$ reduction was utilized in the representation $I^r$. All values in MHz units except where indicated. 
Numbers in parentheses are one standard deviation in units of the least significant figures. 
Empty fields indicate parameters not determined quantum-chemically.
\textsuperscript{a}Ground-state rotational and equilibrium centrifugal distortion parameters. 
\textsuperscript{b}$\rho$, $J_{\rm max}$, $K_{a,{\rm max}}$, the numbers of transitions and lines, and the weighted rms are dimensionless. 
\textsuperscript{c}$\beta$ in units of degrees. 

\label{tab-spec-parameters}
\end{table*}


After each round of measurements and assignments, new spectroscopic parameters were determined. 
The need for additional parameters was tested by investigating the addition of one parameter at a time, 
searching for the one with the largest effect on reducing the weighted rms. 
The calculated sextic centrifugal distortion parameters were added as fixed parameters 
when it appeared that they may affect the lower order parameters. 
They were only released if this resulted in an appreciable reduction of the weighted rms, 
as earlier in the case of the quartic distortion parameters. 
In cases where two parameters provided the greatest improvement in the fit, these two parameters were added 
if the improvement of both parameters combined was approximately equal to the sum of improvement 
contributed by each parameter separately. 
If both parameters combined had about the same effect as one of the two, the one parameter was used 
that appeared to be more in line with the previous parameters. 
In some cases, parameters with magnitudes around five times their uncertainties or less were omitted 
as long as the increase in weighted rms was deemed small enough. 
The resulting spectroscopic parameters are given in Table~\ref{tab-spec-parameters} together with values from 
the B3LYP/3a quantum-chemical calculation.

\section{Discussion}
\label{discussion}
\begin{figure}
\centering
 \includegraphics[width=9cm,angle=0]{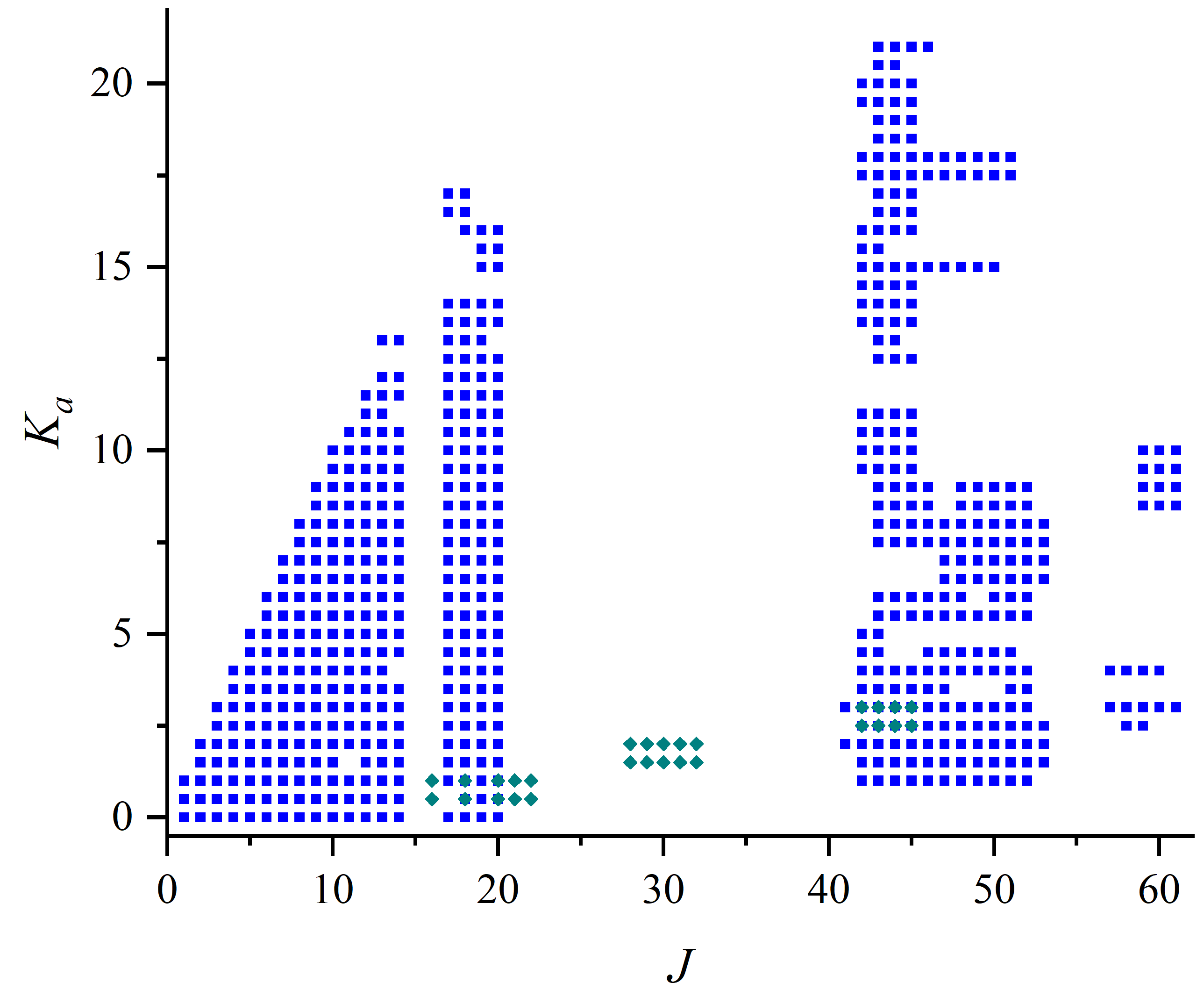}

\caption{Quantum number coverage of the $a$-type transitions in the final fit. 
         Blue squares indicate $R$-branch transitions and green diamonds $Q$-branch transitions. 
         The $K_a = J - K_c$ levels appear as integer values whereas the $K_a = J - K_c +1$ levels 
         appear lowered by 0.5.}
\label{a-type-coverage}
\end{figure}
\begin{figure}
\centering
 \includegraphics[width=9cm,angle=0]{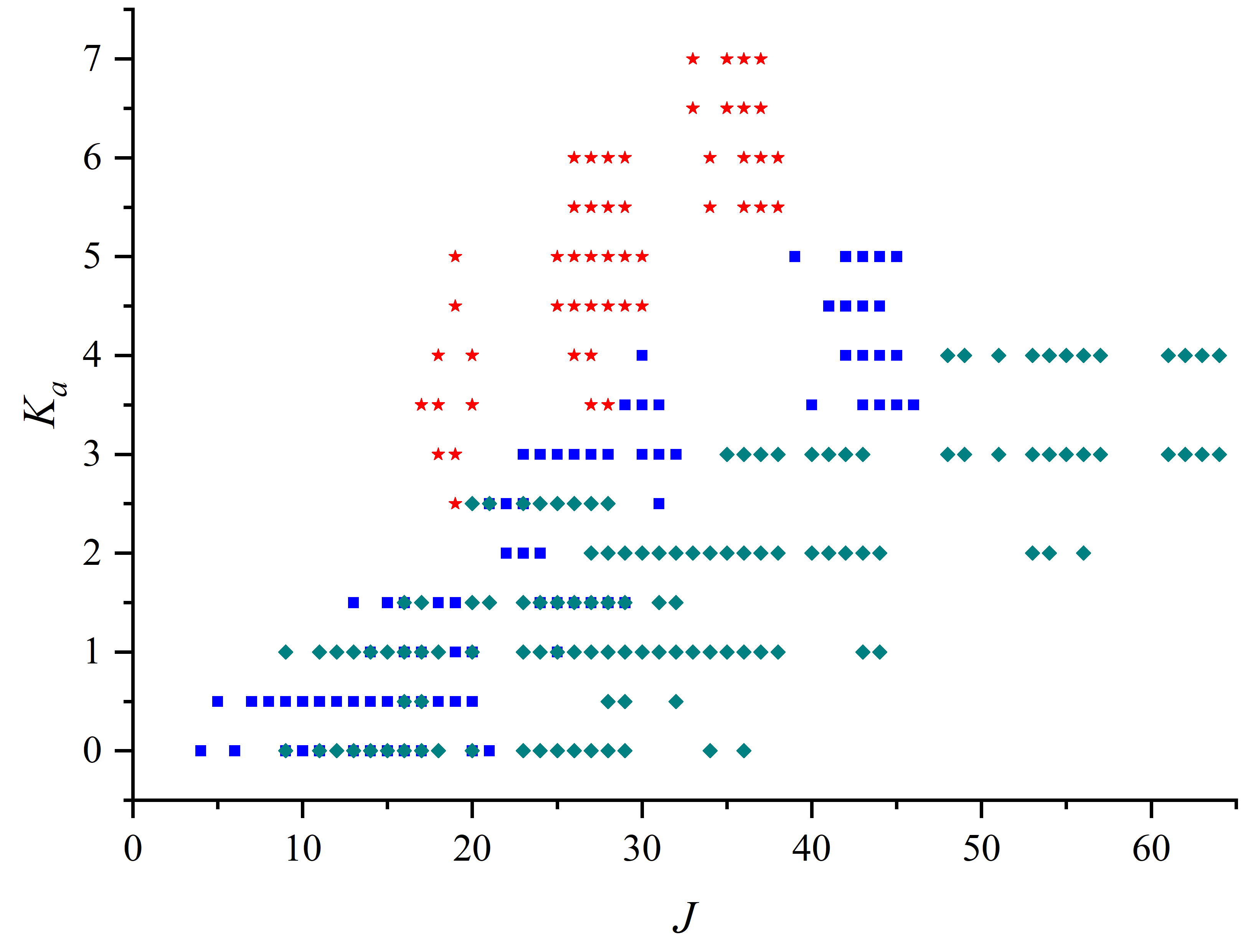}

\caption{Same as Figure~\ref{a-type-coverage}, but for $b$-type transitions. 
         Red stars mark $P$-branch transitions.}
\label{b-type-coverage}
\end{figure}

A total of 961 experimental lines were fit with 33 spectroscopic parameters to within their uncertainties on average. 
This corresponds to 29 lines per parameter on average. 
The quantum number coverage of the $a$-type transitions is shown in Figure~\ref{a-type-coverage}, 
while the one for the $b$-type transitions is shown in Figure~\ref{b-type-coverage}. 
The coverage of the $a$-type $R$-branch transitions reflects the frequency coverage of the present work, 
except that the transitions with $J' \le 4$ are from a previous study \cite{methylallene_rot_1957}. 
And even though the number of $b$-type transitions is much smaller than that of the $a$-types because of their much lower intensity, 
they still cover a fair fraction of the quantum number space, see Figure~\ref{b-type-coverage}.
The lower order as well as some higher order parameter were determined very well, 
that is with uncertainties much smaller than their magnitudes. 
Among the parameters with relatively large uncertainties, most are still well determined 
with their magnitudes being around a factor of 10 larger than their uncertainties. 
Two parameters, $G_{2a}$ and $\epsilon _{JK}$, are not as well determined, but were retained in the fit.

The rotational and centrifugal distortion parameters agree reasonably to well with those calculated at the B3LYP/3a level. 
The method is computationally inexpensive and thus well suited for anharmonic calculations of a somewhat large molecule. 
It is quite common that calculated equilibrium distortion parameters are smaller in magnitude than the experimental ground state values, 
in the present case $A_e = 34845.7$, $B_e = 4201.0$, and $C_e = 3940.7$~MHz.  
However, model insufficiencies are important also as the deviations of a majority of parameters are probably larger than expected for vibrational effects alone. 
Model insufficiencies are most likely also responsible for the modest agreement between experimental dipole moment components,  
$\mu _a = 0.397$~D and $\mu _b = 0.071$~D \cite{methylallene_rot_1957}, compared to 0.457~D and 0.054~D, respectively, from our B3LYP/3a calculation. 
We carried out a MP2 \cite{MPn_1934} calculation with the same triple zeta basis set later which yielded $\mu _a = 0.400$~D and $\mu _b = 0.076$~D, 
in much better agreement with the experimental values.

The Erham program determines a tunneling parameter $\epsilon$. 
Its relation to the CH$_3$ barrier height $V_3$ to internal rotation was investigated empirically in a study on the rotational spectrum of butynone \cite{butynone_rot_2021}. 
The relation displayed some scatter, such that the conversion of one to the other needs to be viewed with some caution. 
Employing $\epsilon = -50.0$~MHz (Table~\ref{tab-spec-parameters}), we obtain $V_3 \approx 513$~cm$^{-1}$, 
in fair agreement with 556~cm$^{-1}$ from a previous rotational study \cite{methylallene_rot_1957}; 
however, this value is based on an assumption on the moment of inertia of CH$_3$ in methylallene. 
On the other hand, their internal rotor splitting value $\Delta _0 = 152.8 \pm 3$~MHz agrees well 
with our splitting value $-3\epsilon = 150.0$~MHz, which is the rotationless A/E splitting.

\section{Conclusion and outlook}
\label{conclusion}

We have recorded and analyzed the rotational spectrum of methylallene into the submillimeter region and obtained accurate spectroscopic parameters. 
The resulting calculated spectrum is accurate enough up to 720~GHz and $J = 90$ with some restrictions concerning high values of $K_a$, 
that is $K_a \gtrsim 20$ for $a$-type transitions and $K_a = 12 - 11$ for $b$-type transitions. 
This is sufficient to search for methylallene not only in cold, dark clouds, but also in the warm parts of star-forming regions in the interstellar medium. 
Calculated spectra were sent to astronomers, however, no unambiguous detection has been made to date. 
Deeper integrations or searches in other sources may eventually lead to its detection.

\section*{Acknowledgement(s)}

We thank the Regionales Rechenzentrum der Universit{\"a}t zu K{\"o}ln (RRZK) for 
providing computing time on the DFG funded High Performance Computing System CHEOPS. 
Our research benefited from NASA's Astrophysics Data System (ADS).

\section*{Disclosure statement}

No potential conflict of interest was reported by the author(s).

\section*{Data availability statement}

A calculated rotational spectrum of the molecule is available in the catalog section of the CDMS 
\cite{CDMS_2001,CDMS_2005,CDMS_2016} at https://cdms.astro.uni-koeln.de/classic/entries/  
as well as in the Virtual Atomic and Molecular Data Centre (VAMDC) \cite{VAMDC_2020} compatible version of the CDMS. 
Line, input, output, and other auxiliary files are available in the data section of the CDMS 
at https://cdms.astro.uni-koeln.de/classic/predictions/daten/Methylallene/. 
All these files have also been deposited at zenodo at https://doi.org/10.5281/zenodo.18376494.

\section*{Funding}

We acknowledge support by the Deutsche Forschungsgemeinschaft (DFG) via the collaborative research center 
SFB~1601 (project ID 500700252) subprojects A4 and Inf as well as the Ger{\"a}tezentrum SCHL~341/15-1 (``Cologne Center for Terahertz Spectroscopy''). 
J.-C. G. is grateful for financial support by the Centre National d'Etudes Spatiales 
(CNES; grant number 4500065585) and by the Programme National Physique et Chimie du 
Milieu Interstellaire (PCMI) of CNRS/INSU with INC/INP co-funded by CEA and CNES.

\section*{ORCID}
Holger S.~P. M\"uller https://orcid.org/0000-0002-0183-8927 \\
Frank Lewen https://orcid.org/0000-0003-4555-2303 \\
Jean-Claude Guillemin https://orcid.org/0000-0002-2929-057X \\
Stephan Schlemmer https://orcid.org/0000-0002-1421-7281


\bibliographystyle{elsarticle-num} 
\bibliography{12BD} 

\end{document}